\documentclass[twocolumn,showpacs,preprintnumbers,amsmath,amssymb]{revtex4}
\usepackage{tipa}
\usepackage{graphicx}
\usepackage{dcolumn}
\usepackage{bm}

\begin{document}

\title{$J^P$ Assignments of $\Lambda_c^+$ Baryons}
\author{Bing Chen$^1$\footnote{chenbing@shu.edu.cn}, Deng-Xia Wang$^1$
and Ailin Zhang$^{1,2}$\footnote{Corresponding author:
zhangal@staff.shu.edu.cn}} \affiliation{$^1$Department of Physics,
Shanghai University, Shanghai 200444, China\\
$^2$Theoretical Physics Center for Science Facilities(TPCSF), CAS,
Beijing 100049, China}

%\date{\today}

\begin{abstract}
The ``good" diquark is employed to study $\Lambda_c^{+}$ baryons within a mass loaded flux tube model. The study indicates that all $\Lambda_c^{+}$ baryons candidates in the 2008 review by the Particle Data Group (PDG) are well described in the mass loaded flux model. The quantum numbers $J^P$ of these $\Lambda_c^{+}$ candidates are assigned. If $\Lambda_c(2765)^{+}$ is an orbitally excited $\Lambda^+_c$, it is likely the $J^P=\frac{3}{2}^{+}$ one. If $\Lambda_c(2765)^{+}$ is an orbitally excited $\Sigma_c$, there ought to be another $J^P=\frac{3}{2}^{+}$ $\Lambda^+_c$ with mass $\approx 2770$ MeV. In the model, there exists no $J^P={1\over 2}^+$ $\Lambda^+_c(\approx 2700)$ predicted in existing literature. $\Lambda_c(2940)^{+}$ is very possible the orbitally excited baryon with $J^P=\frac{5}{2}^{-}$.
\end{abstract}
\maketitle

\section{Introduction}

The study of hadron spectroscopy is an important way to detect the relation between the quark models and QCD, it is also an important way to discover the dynamics in the quark models. The baryon spectroscopy has been explored in many models. Relevant references can be found in some reviews~\cite{review1,review2,review3}. $\Lambda_c^{+}$ baryons have arisen people's great interest for their heavy-light components. In experiments, some $\Lambda_c^{+}$ candidates: $\Lambda^+_c$, $\Lambda_c(2595)^+$, $\Lambda_c(2625)^+$, $\Lambda_c(2765)^+$ (or $\Sigma_c(2765)$), $\Lambda_c(2880)^+$ and $\Lambda_c(2940)^+$ have been observed~\cite{pdg08}. Except for the $J^P$ quantum numbers of $\Lambda_c(2880)^+$, the $J^P$ quantum numbers of most $\Lambda^+_c$ have not been measured. To understand the $\Lambda_c^{+}$ baryons, it is important to pin down the $J^P$ quantum numbers of theses baryons. In this paper, the spectra of $\Lambda_c^{+}$ baryons are studied and the $J^P$ of these $\Lambda^+_c$ are given within a mass loaded flux tube model~\cite{wilczek1,wilczek2,zhang}.

\subsection{Diquark in Baryons}

After the introduction by Gell-Man~\cite{di1}, the diquark was extensively and successfully applied to strong interactions~\cite{di2,di3,di4,di5,di6,di7,di8}. In a modern viewpoint, a diquark is in fact a kind of strong quark correlation in hadrons. When the quark correlations in the wave functions of hadrons are paid attention on, there exist two kinds of diquarks.

The wave function of a diquark consists of
\begin{eqnarray*}
|diquark\rangle=|spatial\rangle\times|color\rangle\times|flavor\rangle\times|spin\rangle.
\end{eqnarray*}

The wave function of the diquark is always assumed to be antisymmetric, and its spacial part and color part are assumed to be symmetric and antisymmetric~\cite{wilczek1}, respectively. Therefore, the wave function
$(|flavor\rangle\times|spin\rangle)_{\emph{diquark}}$ ought to be symmetric. This constraint means that a vector diquark (spin symmetric $3_{\emph{s}}$) is always in flavor symmetric $6_{\emph{f}}$ and a scalar diquark (spin
antisymmetric $\bar{1}_{\emph{s}}$) is always in flavor antisymmetric
$\bar{3}$$_{\emph{f}}$.

In most quark models, the $\emph{color-spin}$ interaction is popularly thought as~\cite{rgg}
\begin{eqnarray*}
\mathcal {H}_{\emph{color-spin}}\sim
-\vec{\sigma}_{i}\vec{\sigma}_{j}\tilde{\lambda}_{i}\tilde{\lambda}_{j}.
\end{eqnarray*}
Such an interaction suggests that the vector diquark has a higher mass than the scalar diquark.
To distinguish these two different kinds of diquarks, the diquark in
($\bar{1}$$_{\emph{s}}$, $\bar{3}$$_{\emph{f}}$) is called a ``good" diquark and denoted as
[$\emph{qq}^{\prime}$], the (3$_{\emph{s}}$, 6$_{\emph{f}}$) diquark is called a ``bad" diquark and denoted as
($\emph{qq}^{\prime}$).

In charmed baryons with two light quarks, the two light quarks are thought to attract each other and be easy to make a diquark~\cite{wilczek1}. It is reasonable to think that there is a ``good" diquark in $\Lambda_{c}$ baryons for their zero isospin ($I=0$). For $\Sigma_{c}$ baryons, there is very possible a ``bad" diquark there for their $I=1$.

In terms of the diquark, the baryon system has the same picture and dynamics as the meson system. The parity of baryons is determined by: $P=(-1)^L$, where $L$ is the orbital angular momentum between the diquark and the third quark.

\subsection{Mass Load Flux Tube Model}

The mass loaded flux tube was studied twenties years ago~\cite{olsson}. Three years ago, this model was explored in terms of the diquark by Selem and Wilczek~\cite{wilczek1}. In the model, the $\Lambda_c^{+}$ baryons are described as follows: the diquark consisting of light u or d quark is connected with the heavy c quark by a flux tube (or a relativistic string) with universal constant tension $\emph{T}$. The difference between the
mass loaded flux model and the valence quark potential model is that the mass loaded flux tube carries both angular momentum and energy~\cite{olsson}.

A S-W formula for the heavy-light systems was given~\cite{wilczek1}
\begin{eqnarray}\label{eq1}
E=M+\sqrt{\frac{\sigma L}{2}}+2^{1/4}\kappa L^{-1/4}m^{3/2}.
\end{eqnarray}

Where the parameter $\sigma$ is the same one as that in the famous
Chew-Frautschi formula $M^{2}=a+\sigma L$. The relationship
between the $\sigma$ and the string tension $\emph{T}$ is
$T=\frac{\sigma}{2\pi}$. The constant $\kappa$ depending on $\sigma$
is defined as: $\kappa\equiv\frac{2\pi^{\frac{1}{2}}}{3\sigma^{\frac{1}{4}}}$.

The spin-orbit interactions were ignored in Eq.~(\ref{eq1}). With the help of this equation, the $J^P$ of $\Lambda_c(2285)^+$, $\Lambda_c(2625)^+$ and $\Lambda_c(2880)^+$ were suggested ${1\over 2}^+$, ${3\over 2}^-$ and ${5\over 2}^+$, respectively, without any analysis~\cite{wilczek1}.

However, as pointed out in Ref.~\cite{zhang}, the spin-orbit interaction may contribute largely to the spectra of $D$ and $D_s$ mesons. Therefore, the spin-orbit interactions ought to be taken into account. A reasonable way to include the contribution of spin-orbit interactions is to simplify the interactions as a $\vec L\cdot \vec S$ coupling or a $\vec{\emph{J}}_{l}\cdot\vec{\emph{J}}_{h}$ coupling, where $J_l$ is the angular momentum of the light quark and $J_h={1\over 2}$ is the angular momentum of the heavy quark. Eq.~(\ref{eq1}) was extended to study the spectra of $D$ and $D_s$ in Ref.~\cite{zhang}.

For $\Lambda_c^{+}$ in the mass loaded flux model, the $\vec{\emph{J}}_{l}\cdot\vec{\emph{J}}_{h}$ coupling is in fact a $\vec{\emph{J}}_{d}\cdot\vec{\emph{J}}_{c}$ coupling, where $J_d$ is the angular momentum of the light diquark. Similarly, Eq.~(\ref{eq1}) is extended to
\begin{eqnarray}\label{eq2}
E=M+\sqrt{\frac{\sigma L}{2}}+2^{1/4}\kappa
L^{-1/4}m^{3/2}+a\vec{J}_{d}\cdot \vec{J}_{c}.
\end{eqnarray}

The parameter $a$ is assumed to depend mainly on heavy flavor and is a near
constant for hadrons with the same heavy flavor. $a$ is determined by experimental data. For the ``good" diquark, $J_d=L$.
%\begin{eqnarray}
%\vec{\emph{J}}_{d}\cdot\vec{\emph{J}}_{c}=\frac{J(J+1)-J_{d}(J_{d}+1)-{3\over 4}}{2}.
%\end{eqnarray}

\section{Analysis of $\Lambda_c^{+}$ baryons}

In the 2008 review by the PDG~\cite{pdg08}, $\Lambda^+_c$, $\Lambda_c(2595)^+$, $\Lambda_c(2625)^+$, $\Lambda_c(2765)^+$ (or $\Sigma_c(2765)$), $\Lambda_c(2880)^+$ and $\Lambda_c(2940)^+$ are listed. However, only the $J^P$ quantum numbers of $\Lambda^+_c(2880)$ have been measured. The $J^P$ of most $\Lambda^+_c$ has never actually been measured, they are assigned according to theoretical expectation. Moreover, the $J^P$ of $\Lambda_c(2765)^+$ (or $\Sigma_c(2765)$) and $\Lambda_c(2940)^+$ have not been given in the review.

$\Lambda_{c}(2765)^{+}$ was first observed in $\Lambda^+_c\pi^+\pi^-$ with a broad width $(\Gamma\approx 50~MeV)$ by
CLEO~\cite{cleo}. However, nothing about its quantum numbers is known. One even does not know whether it is a $\Lambda_c$ or a $\Sigma_{c}$. This state may be a first positive-parity excitation of $\Lambda_c(J^{P}=1/2^{+})$~\cite{cheng}. The $J^P$ of this state was suggested to be ${1\over 2}^+$~\cite{ebert}. This state was suggested to be a $\rho-$mode P-wave excited state with $J^P={1\over 2}^-$~\cite{zhao}. There are some other controversial $J^P$ assignments to it~\cite{other1,other2}

$\Lambda_{c}(2940)^{+}$ was first observed in its decaying into $D^{\circ}P$ by BABAR~\cite{babar} and then confirmed by Belle in $\Sigma_{c}^{0,++}\pi^{+,-}$~\cite{belle} channels. The spin and parity of this state is still unknown. Its $J^P$ was suggested to be ${5\over 2}^-$ or ${3\over 2}^+$~\cite{cheng}. It was supposed to be the D-wave excited state with $J^P={5\over 2}^+$~\cite{zhao}. The radial excitation possibility of this state was excluded~\cite{zhu}. There are also some other controversial interpretations~\cite{other1,other2,other3}.

Obviously, theoretical predictions of $\Lambda_c(2765)^+$ and $\Lambda_c(2940)^+$ is controversial, so the determination of their spin and parity is important. With Eq.~(\ref{eq2}) in hand, we proceed with the study of the spectra of $\Lambda_c^{+}$ baryons. As a byproduct, the $J^P$ of these $\Lambda^+_c$ are suggested.

There are four parameters in Eq.~(\ref{eq2}): mass $M$ of the $c$ quark, mass $m$ of the [$\emph{qq}^{\prime}$] diquark, flux tube tension $\sigma$ and the spin-orbit interaction parameter $a$. To determine these four parameters, a least square fitting process is employed.

In the fitting process, the spin-parity of $\Lambda_{c}(2593)^{+}$ and $\Lambda_{c}(2625)^{+}$ is first set $J^{P}=1/2^{-}$ and $J^{P}=3/2^{-}$, respectively, according to the review by PDG~\cite{pdg08}. Then, different spin-parity attempts of $\Lambda_{c}(2765)^{+}$, $\Lambda_{c}(2880)^{+}$ and $\Lambda_{c}(2940)^{+}$ are made.

Our attempts suggest that $\Lambda_{c}(2765)^{+}$ and $\Lambda_{c}(2880)^{+}$ are the $J^{P}=3/2^{+}$
and the $J^{P}=5/2^{+}$ excited charmed baryons, respectively, and $\Lambda_{c}(2940)^{+}$ is the $J^{P}=5/2^{-}$ excited one. The fitted parameters are: $M_{c}=1.39$ GeV, $m_{d}=0.521$ GeV,
$\sigma=0.999~GeV^2$ and $a=0.04$ GeV. The spectra of $\Lambda_c^{+}$ and their corresponding $J^P$ are given in Table.\ref{table 1}. In the table, all excited $\Lambda_c$ candidates and their possible $J^P$ are put in, and our prediction of the the spectra with above parameters according to Eq.~(\ref{eq2}) are listed.

\begin{table}
\begin{tabular*}{80mm}{c@{\extracolsep{\fill}}cccc}
\toprule L & $J^{P}$(theory)   & $J^{P}$\cite{pdg08} & Mass(theory) & Mass(expt.) \\
\hline
1\hphantom{00} & \hphantom{0}$\frac{1}{2}^{-}$ & \hphantom{0}$\frac{1}{2}^{-}$ & 2584& $\Lambda_{c}(2593)^{+}$ \\
1\hphantom{00} & \hphantom{0}$\frac{3}{2}^{-}$ & \hphantom{0}$\frac{3}{2}^{-}$ & 2644& $\Lambda_{c}(2625)^{+}$ \\
2\hphantom{00} & \hphantom{0}$\frac{3}{2}^{+}$& \hphantom{0}$?^{?}$ & 2772& $\Lambda_{c}^{*}(2765)^{+}$ \\
2\hphantom{00} & \hphantom{0}$\frac{5}{2}^{+}$ & \hphantom{0}$\frac{5}{2}^{+}$ & 2872&$\Lambda_{c}(2880)^{+}$ \\
3\hphantom{00} & \hphantom{0}$\frac{5}{2}^{-}$ & \hphantom{0}$?^{?}$& 2935& $\Lambda_{c}(2940)^{+}$ \\
3\hphantom{00} & \hphantom{0}$\frac{7}{2}^{-}$ & \hphantom{0}$?^{?}$ & 3076& $?$\\
\hline\hline
\end{tabular*}
\caption{Spectrum of excited $\Lambda_c^{+}$ baryons (MeV).}
\label{table 1}
\end{table}

As indicated in Table.\ref{table 1}, the theoretical prediction is perfectly consistent with
the experimental data. In addition, a $J^P={7\over 2}^-$ $\Lambda_c(3076)^+$ is predicted, which is lower than existing theoretical prediction~\cite{isgur,ebert,other2}.

Within a three-quarks system framework, except for the $J^P={1\over 2}^+$ $\Lambda_c(2265)$, another $J^P={1\over 2}^+$ $\Lambda_c(2775)$ was predicted~\cite{isgur}. Similar spectrum pattern of $\Lambda_c$ was predicted in many other models where the baryons are thought as a three-quarks system~\cite{isgur,ebert,other2}. However, in the diquark picture, the expected $J^P={1\over 2}^+$ $\Lambda_c(\approx 2770)$ is a supernumerary state. That is to say, baryons have less resonances in the mass loaded flux tube model than in the three-quarks models.

The mass of the diquark is an important quantity for phenomenological diquark models. In most models, the mass of diquark is $\approx 700$ MeV, it is $521$ MeV in our fitting. The mass of diquark in our analysis seems a little smaller. The exact physics of diquark is not clear and the dynamics in different models may be different, so the mass of diquark may be different. In general, the ``reasonable" mass of the diquark is: $0.35~GeV\sim 0.9~GeV$.

Finally, let us have a look at the mass splittings among these $\Lambda^+_c$. The mass difference among different orbits (with spin-orbit interactions ignored) is also obtained in Table.\ref{table 2}. The experimental data results from a spin average of those observed $\Lambda^+_c$ candidates, and theoretical results come from Eq.~(\ref{eq2}). So far, only one possible $L=3$ excited $\Lambda^+_c$ was observed, therefore experimental $E_{L=3}-E_{L=2}$ and $E_{L=4}-E_{L=3}$ are unknown.
\begin{table}
\begin{tabular*}{80mm}{c@{\extracolsep{\fill}}ccc}
\toprule   & $E_{L=2}-E_{L=1}$ & $E_{L=3}-E_{L=2}$ & $E_{L=4}-E_{L=3}$\\
\hline
expt.\hphantom{0} & \hphantom{0}218 & \hphantom{0}? & ? \\
theory\hphantom{0} & \hphantom{0}209& \hphantom{0}182 & 162  \\
\hline\hline
\end{tabular*}
\caption{Mass difference for different orbits (MeV).}
\label{table 2}
\end{table}

If the excited $\Lambda_{c}^+$ candidates have those assigned $J^P$, the mass splittings resulting from the spin-orbit interactions in the same orbit are obtained (see Table.\ref{table 3}).

\begin{table}
\begin{tabular*}{80mm}{c@{\extracolsep{\fill}}ccc}
\toprule   & L=1 & L=2 & L=3\\
\hline
expt.\hphantom{0} & \hphantom{0}33 & \hphantom{0}115 & ? \\
theory\hphantom{0} & \hphantom{0}60& \hphantom{0}100 & 141  \\
\hline\hline
\end{tabular*}
\caption{Mass splitting in the same orbit (MeV).}
\label{table 3}
\end{table}

Obviously, the spin-orbit interactions are large for higher excited $\Lambda^+_c$. They are even comparable with the splitting among different orbits. Theoretical prediction is more reliable when the spin-orbit interactions has been taken into account.

\section{Discussion and Conclusion}

The diquark has been applied successfully in many phenomenological models. Within the mass loaded flux tube model, the ``good" diquark [$\emph{qq}^{\prime}$] seems to exist in $\Lambda^+_c$ baryons. Our study indicates that excited $\Lambda_c^{+}$ baryons are well described by the mass loaded flux tube. All $\Lambda_c^{+}$ candidates in the review by 2008 PDG have their ``right" place in this model.

Though experimental information about $\Lambda_{c}(2765)^{+}$ is poor, we have some conclusions on this state. If $\Lambda_{c}(2765)^{+}$ is an excited $\Lambda_c^{+}$ baryon, it may be the orbitally excited $(\frac{3}{2})^{+}$ one and not possible the orbitally excited ${1\over 2}^+$ one. If $\Lambda_{c}(2765)^{+}$ is a $\Sigma_{c}$ baryon, there ought to be an orbitally excited $(\frac{3}{2})^{+}$ $\Lambda_c^{+}$ baryon with mass $\approx 2770$ MeV. There exists no $J^P={1\over 2}^+$ $\Lambda^+_c(\approx 2700)$.

$\Lambda_c(2940)^{+}$ is likely the orbitally excited $(\frac{5}{2})^{-}$ $\Lambda^+_c$. In addition, a $(\frac{7}{2})^{-}$ $\Lambda_{c}(3076)^{+}$ is predicted. If this $\Lambda_c(3076)^+$ is observed in the future, it will be a nice test to the mass loaded flux tube model.

Excited $\Sigma_{c}$ baryons are more complex than $\Lambda_c^{+}$ baryons in the dynamics. Experimental data of $\Sigma_{c}$ baryons is poor. How to extend the mass loaded flux tube model to compute the spectra of $\Sigma_{c}$ baryons would be an interesting work. Furthermore, how to study all the heavy-light hadrons in a systematic way in the model deserves further exploration.

Acknowledgment: This work is supported by the National Natural
Science Foundation of China under the grant: 10775093.

\end{document}